\documentclass[preprintnumbers,superscriptaddress,nofootinbib,fleqn,aps,prd]{revtex4}
\pdfoutput=1
\usepackage{hyperref}
\usepackage{amsmath,amssymb,xspace}
\usepackage{graphicx}
\usepackage{subfigure}

\newcommand{\mc}{\mathcal}
\newcommand{\mr}{\mathrm}
\newcommand{\eps}{\varepsilon}

\newcommand{\Dire}{\textsc{Dire}\xspace}
\newcommand{\Pythia}{\textsc{Pythia}\xspace}
\newcommand{\Sherpa}{\textsc{Sherpa}\xspace}

\begin{document}

\preprint{FERMILAB-PUB-21-483-T, LU-TP-21-44, MCNET-21-17}

\title{Disentangling soft and collinear effects in QCD parton showers}
\begin{abstract}
  We introduce a method for the separation of soft and collinear logarithms
  in QCD parton evolution at $\mathcal{O}(\alpha_s^2)$ and at leading color.
  Using an implementation of the technique in the \Dire parton shower, 
  we analyze the numerical impact of genuine triple-collinear corrections
  from quark pair emission in $e^+e^-\to$ hadrons.
\end{abstract}

\author{Leif Gellersen}
\affiliation{Department of Astronomy and Theoretical Physics, Lund University, S-223 62 Lund, Sweden}
\author{Stefan H{\"o}che}
\affiliation{Fermi National Accelerator Laboratory, Batavia, IL, 60510, USA}
\author{Stefan Prestel}
\affiliation{Department of Astronomy and Theoretical Physics, Lund University, S-223 62 Lund, Sweden}

\maketitle

\section{Introduction}
\label{sec:intro}
Monte Carlo event generators have become an indispensable part of the numerical toolkit
needed to interpret high-energy physics experiments at colliders~\cite{Webber:1986mc,Buckley:2011ms}.
They extend the reach of analytic or numeric fixed-order calculations by providing detailed simulations
of QCD parton evolution and hadronization. Both aspects are vital in order to understand the features
of experimentally accessible analysis objects such as jets or photons, and to link
the picture of QCD perturbation theory to the complicated reality of measurements.
Due to the large dynamic range of observables at the Large Hadron Collider (LHC),
the accurate description of QCD evolution plays a particularly important role.
It is implemented in fully differential form by Monte-Carlo algorithms called parton showers.

The high statistical precision of data from the Large Hadron Collider experiments,
as well as the promise of yet more detailed and accurate measurements over the coming years,
have spurred the development of various improved parton shower algorithms.
A number of works have revisited questions on the logarithmic accuracy~\cite{
  Hoche:2017kst,Dasgupta:2018nvj} of parton showers~\cite{Catani:1992ua,Catani:1990rr}
and dipole showers~\cite{Gustafson:1987rq,Lonnblad:1992tz} and have led
to the development of new and improved algorithms~\cite{
  Dasgupta:2020fwr,Bewick:2019rbu,Forshaw:2020wrq,Nagy:2020dvz,Bewick:2021nhc}.
The resummation of logarithms at higher orders in the $1/N_c$ expansion~\cite{
  Platzer:2012np,Nagy:2015hwa,Isaacson:2018zdi,Platzer:2018pmd,Nagy:2019pjp,Forshaw:2019ver,
  Hoche:2020pxj,DeAngelis:2020rvq,Hamilton:2020rcu,Holguin:2020joq,Platzer:2020lbr},
and the possibility to include genuine higher-order matrix elements~\cite{
  Hartgring:2013jma,Li:2016yez,Hoche:2017iem,Dulat:2018vuy} has become a focus
of interest recently. The combination of these various ingredients could soon
enable the formally more precise simulation of QCD parton evolution, and allow
to consistently estimate systematic uncertainties from missing higher-order effects
in the perturbative expansion.

In this note we will focus on the implementation of higher-order splitting kernels
in parton showers. Our numerical implementation is based on a dipole shower, but the
method itself is applicable to any parton shower with on-shell intermediate states.
The possibility of adding next-to-leading order corrections for more inclusive 
observables to parton showers has been explored early on~\cite{
  Kato:1986sg,Kato:1988ii,Kato:1990as,Kato:1991fs,Jadach:2011kc,Gituliar:2014eba}
and was revisited recently~\cite{Hoche:2017hno,Dasgupta:2021hbh}.
A differential approach based on modern shower algorithms was first discussed
in~\cite{Hartgring:2013jma,Li:2016yez}. The link to DGLAP evolution~\cite{
  Gribov:1972ri,Lipatov:1974qm,Dokshitzer:1977sg,Altarelli:1977zs} at next-to-leading
order~\cite{Curci:1980uw,Furmanski:1980cm,Floratos:1980hk,Floratos:1980hm,
  Heinrich:1997kv,Bassetto:1998uv}
was explored in in~\cite{Hoche:2017iem}, and the connection to soft-gluon
resummation~\cite{Korchemsky:1992xv,Korchemsky:1993uz} was established
in~\cite{Dulat:2018vuy}. Here we will address the question of how
higher-order corrections obtained from hard matrix elements in the triple-collinear
and double-soft limits can be combined consistently. 
Our procedure relies on the numerical techniques developed
in~\cite{Hoche:2017iem} and~\cite{Dulat:2018vuy}, which treated the two different limits
individually. We propose a subtraction method that removes soft double
counting at the level of the fully differential evolution kernels for two-parton emission,
and we identify the corresponding endpoint contributions, which are related to the 
two-loop cusp anomalous dimension~\cite{Kodaira:1981nh,Davies:1984hs,Davies:1984sp,Catani:1988vd}.
We apply the method to quark pair emission in the process $e^+e^-\to$ hadrons as an example.

The manuscript is structured as follows: Section~\ref{sec:basic} introduces 
the basic concepts. In Sec.~\ref{sec:tc_ds} we review the techniques 
for the simulation of triple collinear and double soft emissions.
Section~\ref{sec:or} introduces the removal of overlapping singularities,
and Sec.~\ref{sec:mc} presents the modified subtraction needed for a
computation in four dimensions. The endpoint contributions and their relation
to the soft gluon coupling and the CMW scheme~\cite{Catani:1990rr} are
discussed in Sec.~\ref{sec:cmw}.
Section~\ref{sec:results} presents a first numerical analysis,
and Sec.~\ref{sec:conclusion} contains an outlook.

\section{Strategy for constructing an NLO parton shower}
\label{sec:basic}
In this section we provide a heuristic introduction to the main ideas behind a fully differential
parton evolution at next-to-leading order. To this end, it is useful to revisit the basic principles
of a leading-order algorithm.

The one-loop matrix elements for gluon emissions off a color dipole exhibit two types
of singularities~\cite{Ellis:1991qj,Dokshitzer:1991wu}: soft gluon singularities and collinear poles.
Most of the existing leading-order parton shower algorithms treat these two effects in a unified way:
They either employ one splitting kernel to describe the complete antenna radiation pattern,
or two splitting kernels that capture the collinear monopole radiation pattern. In the first case,
the collinear radiator function is matched to the soft, while in the second case the soft radiator
function is matched to the collinear, while at the same time removing potential double counting 
through partial fractioning of eikonal terms or angular ordering.

To construct a parton shower at next-to-leading order
accuracy, it is useful to discard this picture and instead recall that the soft gluon limit
has a semi-classical origin and is thus structurally different from the collinear limit.
However, the two do of course overlap in the soft-collinear region.
An improved leading-order parton-shower can therefore be constructed by working with 
three different radiator functions for each color dipole, one capturing the soft emission
pattern, and one each for capturing the remainder of the collinear radiators, after subtracting
the overlap with the soft function. This strategy allows to cover the complete phase space
with each evolution kernel, and it furthermore allows to choose different evolution
variables in the soft and collinear regions.
Representative squared diagrams for a process with two hard partons are
\begin{equation}
    \label{eq:lops}
    \left.\mathcal{F}\right|_\mathrm{1-loop,coll} \sim
    \vcenter{\hbox{\includegraphics[width=.15\textwidth]{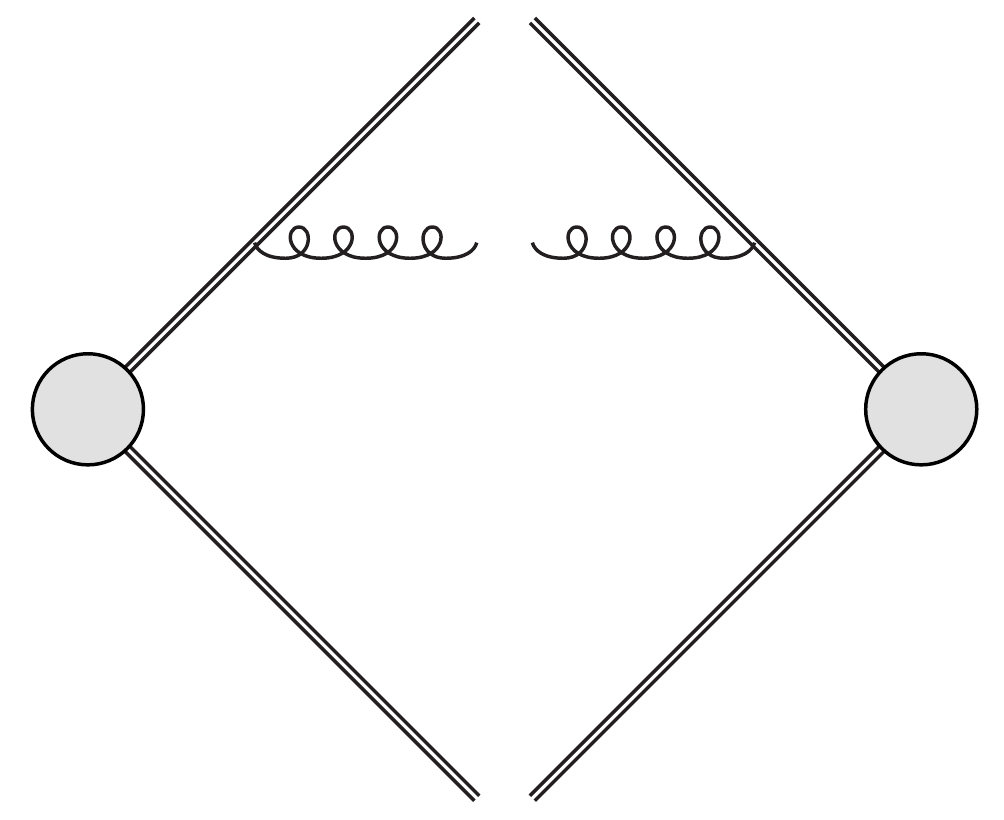}}}\;,\qquad
    \left.\mathcal{F}\right|_\mathrm{1-loop,soft} \sim
    \vcenter{\hbox{\includegraphics[width=.15\textwidth]{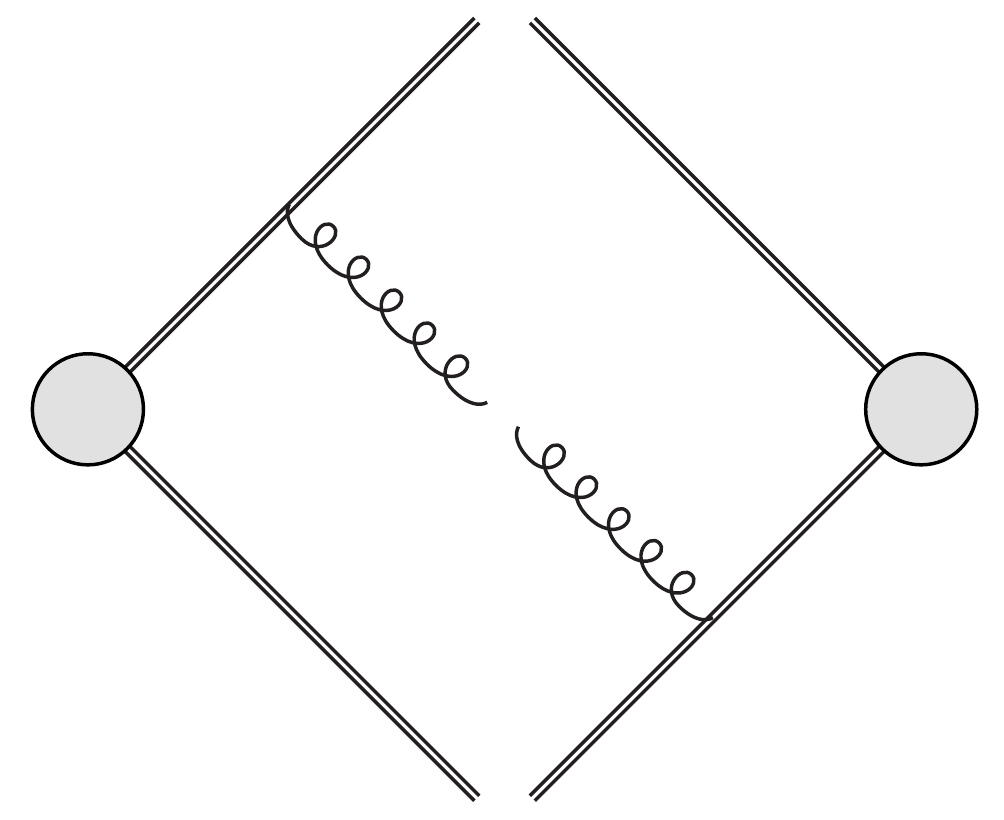}}}
    ~+~\ldots\;,
\end{equation}
where the dots stand for diagrams with permutations of the hard partons. The left figure 
indicates a collinear emission, and the right indicates coherent soft gluon radiation.
At second order in the strong coupling, the perturbative fragmentation functions
will contain real-virtual and double-real corrections. We will use the emission 
of a quark pair as an example for the construction of a soft-collinear overlap removal
in these contributions. The triple-collinear $q\to qq'\bar{q}'$ splitting function
can be factorized into a collinear one-loop $q\to g$ times a collinear one-loop $g\to q$ 
splitting in the strongly ordered limit, while the soft function for quark-pair emission
cannot be factorized into lower-order soft functions. However, it can be factorized 
into a product of eikonal currents times a spin-dependent collinear one-loop $g\to q$ splitting,
in fact it is given entirely in terms of their product~\cite{Catani:1999ss}. Effective diagrams 
for double-real corrections at two loops may thus be approximated by iterated branchings, 
\begin{equation}
    \label{eq:lops_oas2_fac}
    \left.\mathcal{F}\right|_{\mathrm{2-loop,coll}}
    \sim\vcenter{\hbox{\includegraphics[width=.133\textwidth]{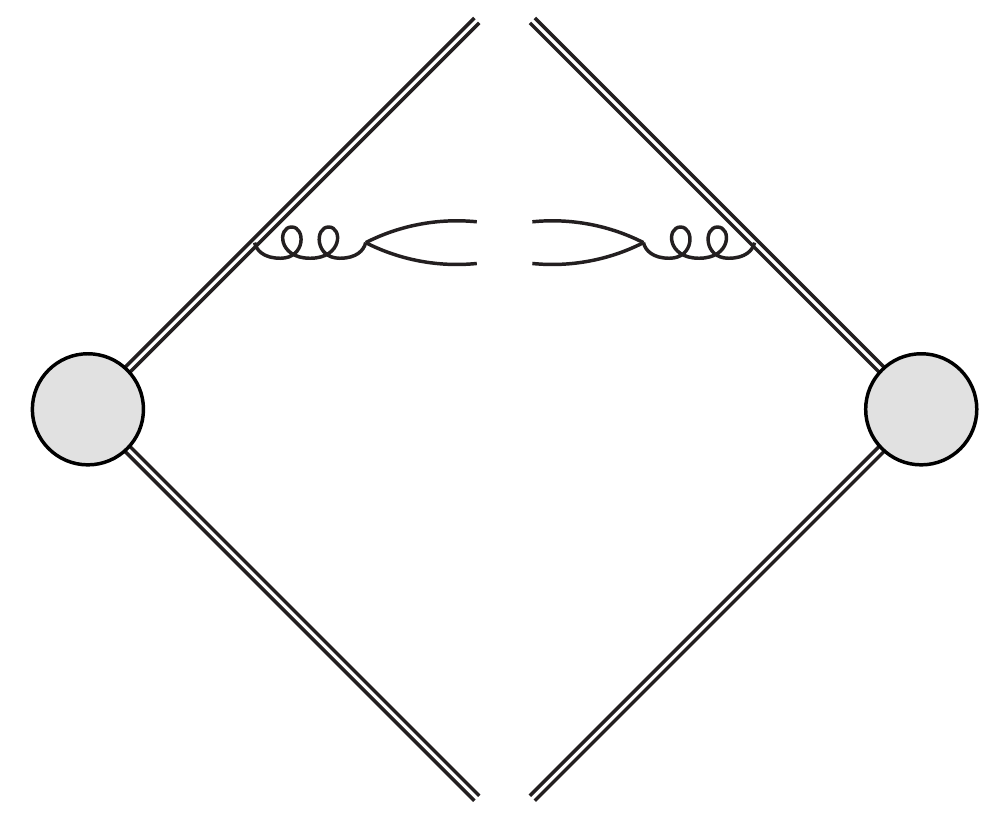}}}\;,
    \qquad\left.\mathcal{F}\right|_{\mathrm{2-loop,soft}}
    \sim\vcenter{\hbox{\includegraphics[width=.133\textwidth]{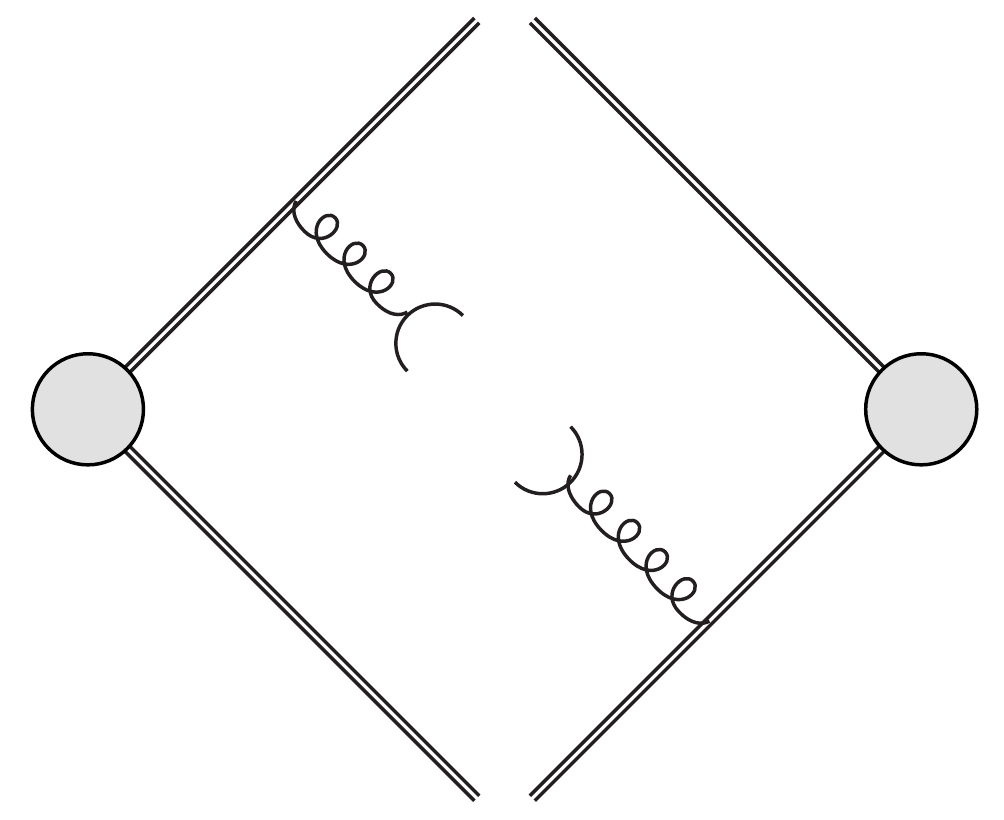}}}
    ~+~\ldots\;.
\end{equation}
In analogy to fixed-order computations in the dipole method, the calculation of the double-real 
corrections to this approximate picture proceeds by subtracting the approximate result in
Eq.~\eqref{eq:lops_oas2_fac} from the complete matrix elements. In addition, an endpoint contribution 
is required, which originates in the difference between the integrated subtraction terms
and the corresponding collinear mass factorization counterterms. The result is finite 
in four dimensions and can therefore be computed with Monte-Carlo methods~\cite{Hoche:2017iem}.
Using the double-real quark-pair emission triple-collinear (tc) and double-soft (ds) kernels 
as an example, we can write, schematically
\begin{equation}
    \label{eq:pds}
    P^{(\mathrm{tc})} \sim  \left[
    \vcenter{\hbox{\includegraphics[width=.133\textwidth]{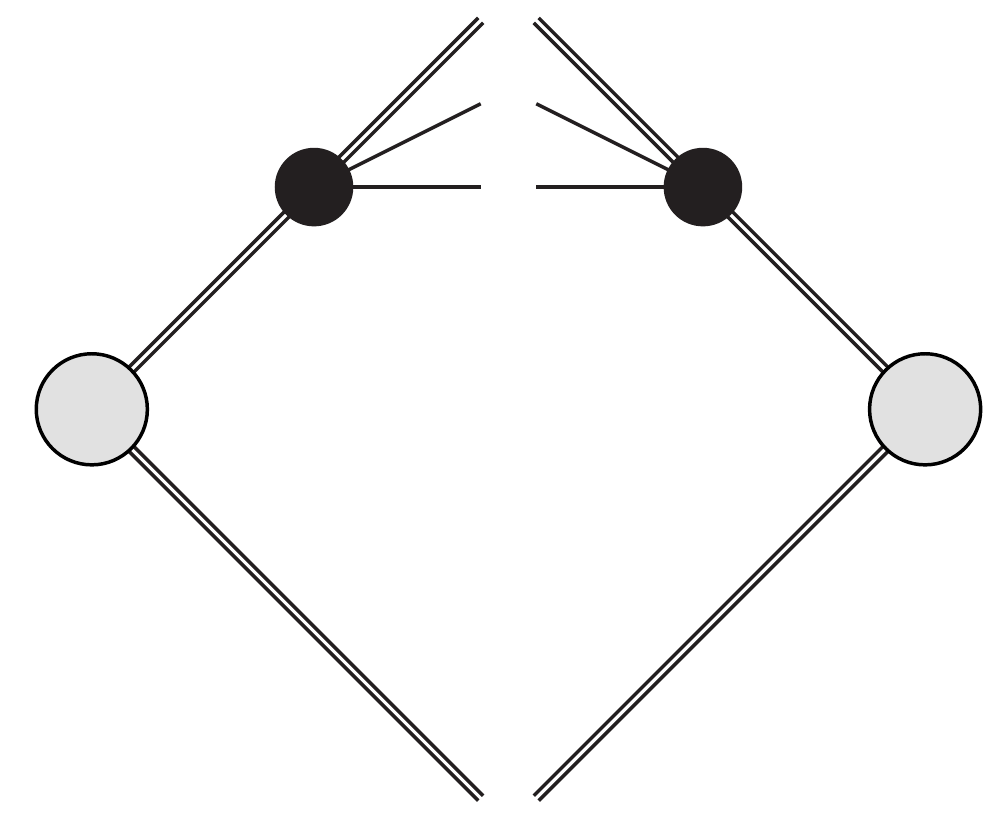}}}
    ~-~\vcenter{\hbox{\includegraphics[width=.133\textwidth]{fig/intro/tcds-qqqq-rrlo-sc-1}}}\right]\;,
    \qquad
    P^{(\mathrm{ds})} \sim \left[
    \vcenter{\hbox{\includegraphics[width=.133\textwidth]{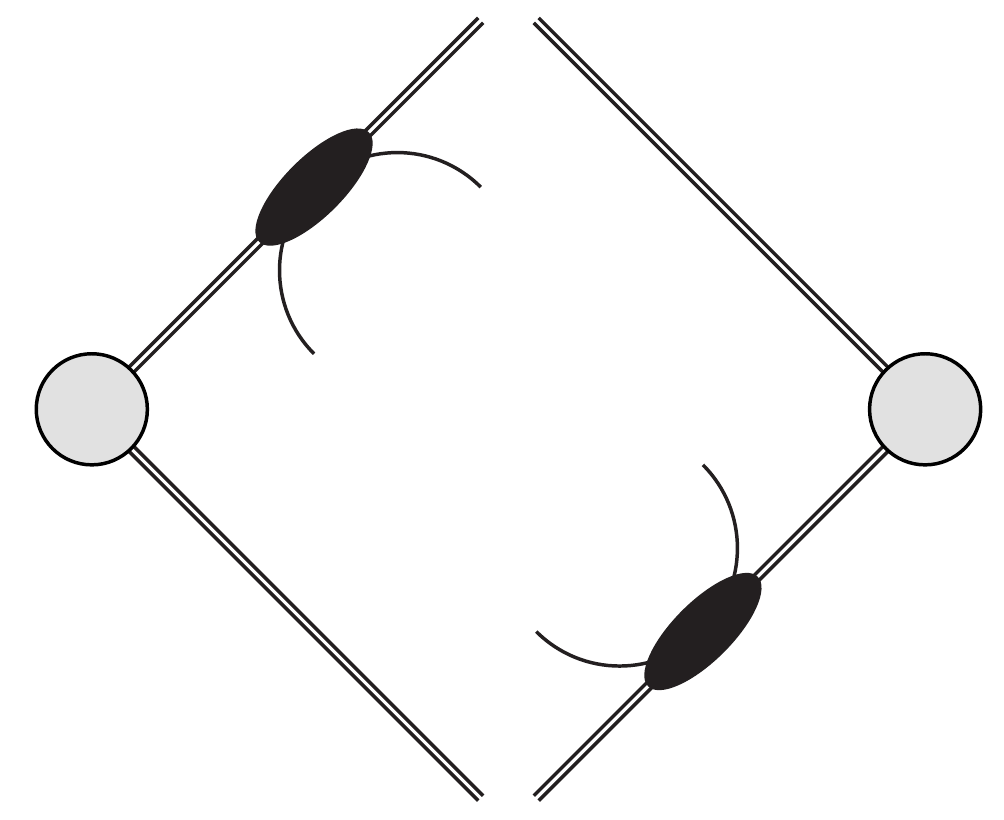}}}
    ~-~\vcenter{\hbox{\includegraphics[width=.133\textwidth]{fig/intro/tcds-qqqq-rrlo-ss-1}}}
    ~+~\ldots~\right]\;,
\end{equation}
where the black blobs indicate the complete matrix elements in the triple collinear and double soft limits.
Equation~\eqref{eq:pds} is valid independently for both the differential and the endpoint contributions.
For an appropriately defined leading-order parton shower, this subtraction must remove all
infrared singularities associated with the vanishing of intermediate propagators. This puts
stringent requirements on the leading-order shower, in particular that it must implement spin 
correlations and a suitable kinematics mapping~\cite{Hoche:2017iem,Dulat:2018vuy}.

The above subtraction ensures that the correct splitting probabilities are reproduced in
the collinear and soft region individually, but it is insufficient to guarantee the correct 
two-loop radiation pattern in multiple limits simultaneously, because the individual two-loop 
splitting functions have overlapping singularities. Each triple-collinear matrix element
contains the complete double-soft result. This is reminiscent of the overlap of the double
collinear and single soft matrix elements in the leading-order case. To remove the overlap,
a solution similar to the leading-order case can be adopted: A combination of 
triple-collinear and double-soft corrections at leading color requires 1) 
removing the endpoint-subtracted double-soft splitting function from the 
endpoint-subtracted triple-collinear splitting function, and
2) adding the double-soft splitting functions for all pairs of hard partons and
the soft-subtracted collinear splitting functions for all partons in order to obtain 
the complete radiator function for the multipole. In the case of quark pair emission, 
the genuine triple collinear contributions to this combined splitting function are given by
\begin{eqnarray}
    \label{eq:ptcds}
    P^{(\mathrm{tc}-\mathrm{ds})} &\sim& \left[
    \vcenter{\hbox{\includegraphics[width=.133\textwidth]{fig/intro/tcds-qqqq-rrtc-1}}}
    ~-~\vcenter{\hbox{\includegraphics[width=.133\textwidth]{fig/intro/tcds-qqqq-rrlo-sc-1}}}
    ~-~\vcenter{\hbox{\includegraphics[width=.133\textwidth]{fig/intro/tcds-qqqq-rrds-1}}}
    ~+~\vcenter{\hbox{\includegraphics[width=.133\textwidth]{fig/intro/tcds-qqqq-rrlo-ss-1}}}
    ~+~\ldots~\right]\;.
\end{eqnarray}
Again, this is valid independently for both the differential and the endpoint contributions.
The subtraction has to be applied for every possible occurrence of the double-soft limit
in the triple-collinear splitting functions. 
In the following sections, we will first discuss the individual triple collinear
and double soft limits of the QCD matrix elements, and then develop the above
described procedure in detail for quark pair emission. The gluon emission case 
is structurally identical but technically more involved. We postpone its discussion
to a forthcoming publication.

\section{Parton evolution in the triple collinear and double soft limits}
\label{sec:tc_ds}
In this section we summarize the ingredients needed for the consistent simulation
of triple collinear and double soft splittings in a dipole-like parton shower.
We note that this type of parton shower is affected by the problems discussed
in~\cite{Dasgupta:2018nvj}, but the structure of our calculation is generic
and can therefore be applied to any parton shower for which the phase-space
factorization and splitting functions are known in $D=4-2\eps$ dimensions.

In the triple collinear limit of partons $1$, $2$ and $3$, any QCD associated
matrix element with more than 3 external partons factorizes as~\cite{Campbell:1997hg,Catani:1999ss}
\begin{equation}\label{eq:tc_me_factorization}
  |M_{1,2,3,\ldots,k,\ldots}(p_1,p_2,p_3,\ldots)|^2\overset{\rm 123-coll}{\longrightarrow}
  \left(\frac{8\pi\mu^{2\eps}\alpha_s}{s_{123}}\right)^2
  \mc{T}^{ss'}_{123,\ldots}(p_{123},\ldots)\,
  P^{ss'}_{123}(p_1,p_2,p_3)\;.
\end{equation}
The corresponding spin-averaged, triple-collinear splitting functions, 
$\delta_{ss'}P^{ss'}_{123}/2$, are given in~\cite{Campbell:1997hg,Catani:1999ss}. 
The simplest of them are the quark to quark splitting kernels with quark pair emission. 
They read
\begin{equation}\label{eq:p_qbpqpq}
  \begin{split}
    P_{\bar{q}_1'q_2'q_3}=&\;\frac{1}{2}C_FT_R\frac{s_{123}}{s_{12}}
    \left[\frac{4z_3+(z_1-z_2)^2}{z_1+z_2}-\frac{t_{12,3}^2}{s_{12}s_{123}}
      +(1-2\eps)\left(z_1+z_2-\frac{s_{12}}{s_{123}}\right)\right]\;,\\
    P_{\bar{q}_1q_2q_3}=&\;\Big[\,P_{\bar{q}_1'q_2'q_3}+P_{\bar{q}_1'q_3'q_2}\,\Big]
    +\Big[\,P_{\bar{q}_1q_2q_3}^{\rm(id)}+P_{\bar{q}_1q_3q_2}^{\rm(id)}\,\Big]\;,
  \end{split}
\end{equation}
where $s_{ij}=2p_ip_j$ are the scalar products of the (light-like) parton momenta,
$s_{123}=s_{12}+s_{13}+s_{23}$, and where $z_i=p_in/p_{123}n$ is the light-cone momentum fraction 
of particle $i$ with respect to an arbitrary auxiliary vector $n$, which must not be parallel
to the collinear momentum, $p_{123}=p_1+p_2+p_3$.
The interference term, $P_{\bar{q}_1q_2q_3}^{\rm(id)}$, is given by
\begin{equation}\label{eq:p_qbqq_id}
  \begin{split}
    P_{\bar{q}_1q_2q_3}^{\rm(id)}=&\;C_F\left(C_F-\frac{C_A}{2}\right)\bigg\{
    (1-\eps)\left(\frac{2s_{23}}{s_{12}}-\eps\right)
    -\frac{s_{123}^2}{s_{12}s_{13}}\frac{z_1}{2}\left[\frac{1+z_1^2}{(1-z_2)(1-z_3)}
      -\eps\left(1+2\frac{1-z_2}{1-z_3}\right)-\eps^2\right]\\
    &\qquad+\frac{s_{123}}{s_{12}}\left[\frac{1+z_1^2}{1-z_2}-\frac{2z_2}{1-z_3}
      -\eps\left(\frac{(1-z_3)^2}{1-z_2}+1+z_1-\frac{2z_2}{1-z_3}\right)-\eps^2(1-z_3)\right]\bigg\}\;.
  \end{split}
\end{equation}
Following~\cite{Catani:1999ss}, we have defined
\begin{equation}\label{eq:t123}
  t_{12,3}=2\,\frac{z_1s_{23}-z_2s_{13}}{z_1+z_2}+\frac{z_1-z_2}{z_1+z_2}\,s_{12}\;.
\end{equation}
We can interpret the triple collinear branching of the combined parton $(123)$
as two subsequent splittings, $(123)\to(12)3$ and $(12)\to12$. Integration over
the final-state phase space of the second splitting, renormalization and collinear
mass factorization in the $\overline{\rm MS}$ scheme then lead to the integrated
double-collinear time-like splitting functions at NLO accuracy~\cite{
  Curci:1980uw,Furmanski:1980cm,Floratos:1980hk,Floratos:1980hm,Heinrich:1997kv,Bassetto:1998uv}
\begin{equation}\label{eq:p1_qqp}
  \begin{split}
    P_{qq'}^{(T)}(z)=&\;C_F T_R\left((1+z)\log^2(z)
      -\left(\frac{8}{3}z^2+9z+5\right)\log(z)+\frac{56}{9}z^2+4z-8-\frac{20}{9z}\right)\;,\\
    P_{q\bar{q}}^{(T)}(z)=&\;P_{qq'}^{(T)}(z)+C_F \bigg(C_F-\frac{C_A}{2}\bigg)\bigg(2p_{qq}(-z)S_2(z)
      +2(1+z)\log(z)+4(1-z)\bigg)\;,\\
  \end{split}
\end{equation}
where $p_{qq}(z)=(1+z^2)/(1-z)$, and where the auxiliary function $S_2$ is defined as
\begin{equation}
    S_2(z)=-2{\rm Li}_2\frac{1}{1+z}+\frac{1}{2}\ln^2z-\ln ^2(1-z)+\frac{\pi^2}{6}\;.
\end{equation}
In the double soft limit, the hard matrix element for emission of a quark-antiquark pair factorizes
as~\cite{Campbell:1997hg,Catani:1999ss}
\begin{equation}\label{eq:soft_factorization}
    |M_{1,2,3,\ldots,n}(p_1,p_2,p_3,\ldots,p_n)|^2
    \overset{\rm 12-soft}{\longrightarrow}\left(4\pi\mu^{2\eps}\alpha_s\right)^2
    \sum_{\substack{i,j=3}}^{n}\mathcal{I}_{ij}(p_1,p_2)\,|M_{3,\ldots,n}^{(i,j)}(p_3,\ldots,p_n)|^2,
\end{equation}
where the color-correlated tree-level matrix element squared is given by
\begin{equation}\label{eq:color_correlated_born}
    |M_{3,\ldots,n}^{(i,j)}(p_3,\ldots,p_n)|^2=
    -\langle M_{3,\ldots,n}(p_3,\ldots,p_n)|\,\hat{T}_i\hat{T}_j\,|M_{3,\ldots,n}(p_3,\ldots,p_n)\rangle\;.
\end{equation}
The corresponding double-soft splitting function, $\mathcal{I}_{ij}(p_1,p_2)$,
is given by~\cite{Campbell:1997hg,Catani:1999ss}
\begin{equation}\label{eq:ds_qbpqpq}
  \mathcal{I}_{ij}(p_1,p_2)=T_R\,
  \frac{s_{i1}s_{j2}+s_{i2}s_{j1}-s_{ij}s_{12}}{
    s_{12}^2(s_{i1}+s_{i2})(s_{j1}+s_{j2})}\;.
\end{equation}
In contrast to the one-loop case, the $i=j$ contributions to the
soft matrix element in Eq.~\eqref{eq:soft_factorization} do not vanish. 
In the following section, we will discuss the combination of
Eqs.~\eqref{eq:tc_me_factorization} and~\eqref{eq:soft_factorization}
in a fully differential parton-shower simulation.

\section{Overlap removal and genuine collinear anomalous dimension}
\label{sec:or}
Following the general arguments outlined in Sec.~\ref{sec:basic}, we need to remove the 
collinear limit of the double-soft matrix element, Eq.~\eqref{eq:soft_factorization}, 
from the triple-collinear matrix element, Eq.~\eqref{eq:tc_me_factorization}, in order
to obtain a purely collinear remainder. In this limit, we can perform the sum over
spectator partons, $j$, in Eq.~\eqref{eq:soft_factorization}, while holding 
$i=3$ fixed. This yields the collinear limit of the soft factorization formula
\begin{equation}\label{eq:soft_factorization_coll}
    |M_{1,2,3,\ldots,n}(p_1,p_1,p_3,\ldots,p_n)|^2
    \overset{\rm 12-soft}{\underset{\rm 123-coll}{\longrightarrow}}
    \left(\frac{8\pi\mu^{2\eps}\alpha_s}{s_{123}}\right)^2
    \mathcal{T}^{ss}_{123,\ldots}(p_{123},\ldots)P^{\rm(ds)}_{123}(p_1,p_2,p_3)\;,
\end{equation}
where the double soft splitting function, $P^{\rm(ds)}_{123}$, is given
by~\cite{Dulat:2018vuy}
\begin{equation}\label{eq:ps_qbpqpq}
  P_{\bar{q}_1'q_2'a_3}^{\rm(ds)}=\frac{1}{2}C_aT_R\,\frac{s_{123}^2}{(s_{13}+s_{23})^2}
  \left[\frac{4z_3}{1-z_3}\frac{s_{13}+s_{23}}{s_{12}}
    -\left(\frac{t_{12,3}}{s_{12}}-\frac{z_1-z_2}{z_1+z_2}\right)^2\,\right]\; . \\
\end{equation}
This function can be integrated using the phase-space parametrization
of~\cite{Gehrmann-DeRidder:2003pne}. Following~\cite{Hoche:2017iem}, we 
factor out the two-particle phase space, the integration over the three-particle invariant
$y_{aij}=s_{aij}/q^2$ and the corresponding factors $(y_{aij}(1-y_{aij}))^{1-2\eps}$
as well as the integration over one of the light-cone momentum fractions,
which is chosen to be $\tilde{z}=s_{ak}/q^2/(1-y_{aij})$. We also remove the square
of the normalization factor $(4\pi)^{\eps}/(16\pi^2\Gamma(1-\eps))\,(q^2)^{1-\eps}$.
The remaining one-emission phase-space integral reads
\begin{equation}\label{eq:tcps_tl}
  \begin{split}
    \int{\rm d}\Phi_{+1}^{(F)}=&\;
    (1-\tilde{z})^{1-2\eps}\tilde{z}^{-\eps}
    \int_0^1{\rm d}\tau\,(\tau(1-\tau))^{-\eps}
    \int_0^1{\rm d}v\,(v(1-v))^{-\eps}\;
    \frac{\Omega(1-2\eps)}{\Omega(2-2\eps)}
    \int_0^1{\rm d}\chi\,2(4\chi(1-\chi))^{-1/2-\eps}\;,
  \end{split}
\end{equation}
where $\Omega(n)=2\pi^{n/2}/\Gamma(n/2)$.
The variables $\tau$ and $v$ are given by the transformation \cite{Hoche:2017iem}
\begin{equation}
  s_{ai}=s_{aij}(1-\tilde{z}_j)\,v\;,
  \qquad
  \tilde{z}_j=\frac{s_{jk}/q^2}{1-y_{aij}}=(1-\tilde{z})\,\tau\;.
\end{equation}
The azimuthal angle integration is parametrized using $\chi$, which is defined as
$ 
  s_{ij}=s_{ij,-}+\chi(s_{ij,+}-s_{ij,-})\,,
$ 
with $s_{ij,\pm}$ being the two solutions of the quadratic equation
$\cos^2\phi_{a,i}^{j,k}=1$~\cite{Gehrmann-DeRidder:2003pne}.
The result is
\begin{equation}
    \begin{split}
    &\frac{1}{C_aT_R}\int{\rm d}\Phi_{+1}^{(F)}P_{aq'}^{(ds)}
    =-\frac{1}{\eps}\left(\frac{4}{3\tilde{z}}-2\tilde{z}+\frac{2\tilde{z}^2}{3}+2\ln\tilde{z}\right)\\
    &\qquad-2\left({\rm Li}_2(\tilde{z})-\zeta_2\right)+3\ln^2\tilde{z}
    +\frac{2}{3\tilde{z}}(1-7\tilde{z}+10\tilde{z}^2-4\tilde{z}^3)\\
    &\qquad+\left(\frac{8}{3\tilde{z}}-2\tilde{z}+\frac{2\tilde{z}^2}{3}\right)\ln\tilde{z}
    +\left(\frac{4}{3\tilde{z}}-2\tilde{z}+\frac{2\tilde{z}^2}{3}\right)\ln(1-\tilde{z})
    +\mathcal{O}(\eps) \, .
    \end{split}
\end{equation}
Upon including the propagator term from Eq.~\eqref{eq:soft_factorization_coll} and
the phase-space factor $y_{aij}^{1-2\eps}$, the leading pole is multiplied by an additional
factor $-\delta(y_{aij})/2\eps$. The $1/\eps^2$ coefficient thus generated is removed
by the renormalization of the soft component of the fragmentation function. 
This renormalization term is obtained as
\begin{equation}\label{eq:ffpdf_ren}
  \mc{P}_{aq'}^{(ds)}(\tilde{z})=\int_{\tilde{z}}^1\frac{{\rm d}x}{x}
  P_{ag}^{\rm(0,s)}(z)P_{gq}^{(0)}(z/x)=\,C_a T_R\left(2\ln\tilde{z}
  +\frac{2\tilde{z}^2}{3}-2\tilde{z}+\frac{4}{3\tilde{z}}\right)\;,
\end{equation}
where $P_{ag}^{\rm(0,s)}(z)=2C_a(1-z)/z$ is the soft limit of the double-collinear
splitting function for gluon emission.
In order to extract the analog of the next-to-leading order splitting function $P_{qq'}$,
we employ the two-loop matching condition for the fragmenting jet function~\cite{Ritzmann:2014mka}.
\begin{equation}\label{eq:twoloop_ffmatch}
  \begin{split}
    \mc{G}_a^{i(2)}(s,z,\mu)=\mc{J}_{ai}^{(2)}(s,z,\mu)                                                             
    +\sum_j\int_z^1\frac{{\rm d}x}{x}\mc{J}_{aj}^{(1)}(s,z/x,\mu)D_j^{i(1)}(x,\mu)                                  
    +\delta(s)D_a^{i(2)}(z,\mu)\;.                                                                                  
  \end{split}                                                                                                       
\end{equation}
The complete matching term is given in~\cite{Ritzmann:2014mka,Hoche:2017iem}. 
Its soft-collinear analog needed for $\mc{G}_a^{q'(2)}$ is given by
\begin{equation}\label{eq:match_tl}
  \int_z^1\frac{{\rm d}x}{x}\mc{J}_{ag}^{(1)}(s,z/x,\mu)D_g^{q(1)}(x,\mu)\,\Big|_{\rm(ds)}
  =2\int_{\tilde{z}}^1\frac{{\rm d}x}{x}
  \,2C_F\,\frac{1-x}{x}\ln(x(1-x))\;P_{gq}^{(0)}(\tilde{z}/x)\;.
\end{equation}
A detailed discussion will be given in Sec.~\ref{sec:cmw}.
Using this technique, we obtain the soft-collinear contribution to the
timelike NLO $q\to q'$ splitting function
\begin{equation}\label{eq:p1_qqp_ds}
    \begin{split}
    P_{qq'}^{\rm(ds,T)}(\tilde{z})=&\;
    C_a T_R\bigg(2({\rm Li}_2\tilde{z}-\zeta_2)+\ln^2\tilde{z}
    -\left(8+6\tilde{z}-2\tilde{z}^2\right)\ln \tilde{z}\\
    &\qquad\qquad\left.-\left(\frac{4}{3\tilde{z}}-2\tilde{z}+\frac{2\tilde{z}^2}{3}\right)\ln(1-\tilde{z})
    -\frac{34}{9}\tilde{z}^2+\frac{46}{3}\tilde{z}-\frac{28}{3}-\frac{20}{9\tilde{z}}\right)\;.
    \end{split}
\end{equation}
In the numerical simulation this term can be obtained from the quark-pair contribution 
to the double soft splitting function, taken in the triple collinear limit.
The corresponding methods have been discussed in detail in~\cite{Dulat:2018vuy}, 
and here we will therefore focus on the difference to the complete timelike NLO $q\to q'$ 
splitting function only. This difference leads to a genuine two-loop timelike collinear 
anomalous dimension that is given by
\begin{equation}
    \gamma_{qq'}^{\rm(tc,T)}=\int_0^1{\rm d}z\,z\left(P_{qq'}^{\rm(T)}(z)-P_{qq'}^{\rm(ds,T)}(z)\right)=
    \left(\frac{11}{18}-\frac{2\pi^2}{9}\right)C_FT_R\;. \label{eq:collAnomDim}
\end{equation}
Based on this result, we expect the genuinely triple collinear configurations to generate
a small negative correction to the leading-order radiation pattern. 
In the following section we will discuss how the above computation can be implemented
using a four-dimensional modified subtraction scheme.

\section{Computation in four dimensions}
\label{sec:mc}
The modified subtraction procedure needed to implement next-to-leading order corrections to the
parton-shower splitting kernels was outlined in~\cite{Hoche:2017iem}. The computation of the soft 
contributions is performed according to the formula
\begin{equation}\label{eq:tcps_mc}
  P_{aq'}^{\rm(ds)}(\tilde{z})=\Big(\mr{I}+\frac{1}{\eps}\,\mc{P}-\mc{I}\Big)_{aq'}^{\rm(ds)}(\tilde{z})+
  \int{\rm d}\Phi_{+1}(\mr{R}-\mr{S})_{aq'}^{\rm(ds)}(\tilde{z},\Phi_{+1})\;. 
\end{equation}
In order to implement the algorithm, we need the approximate spin-independent splitting function,
$\tilde{P}_{aq'}^{1\to3\rm(ds)}$ and the corresponding spin correlation term, 
$\Delta\tilde{P}_{aq'}^{1\to3\rm(ds)}$, which define the differential subtraction term according to
\begin{equation}\label{eq:dsps_def_rs}
  \begin{split}
    \mr{R}_{aq'}^{\rm(ds)}(\tilde{z},\Phi_{+1})=&\;P_{aq'}^{1\to3\rm(ds)}(\tilde{z},\Phi_{+1})\\
    \mr{S}_{aq'}^{\rm(ds)}(\tilde{z},\Phi_{+1})=&\;\tilde{P}_{aq'}^{1\to3\rm(ds)}(\tilde{z},\Phi_{+1})
    +\Delta\tilde{P}_{aq'}^{1\to3\rm(ds)}(\tilde{z},\Phi_{+1})\;.
  \end{split}
\end{equation}
The two contributions to the subtraction term are given by
\begin{equation}\label{eq:tcsf_qqpa}
  \begin{split}
  \tilde{P}_{aq'}^{1\to3\rm(ds)}(\tilde{z}_a,\tilde{z}_i,\tilde{z}_j,s_{ai},s_{aj},s_{ij})
  =&\;C_a T_R\frac{s_{aij}}{s_{ai}}\frac{2\tilde{z}_j}{1-\tilde{z}_j}
  \left(1-\frac{2}{1-\eps}\frac{\tilde{z}_a\tilde{z}_i}{(\tilde{z}_a+\tilde{z}_i)^2}\right)\\
  \Delta\tilde{P}_{aq'}^{1\to3\rm(ds)}(\tilde{z}_a,\tilde{z}_i,\tilde{z}_j,s_{ai},s_{aj},s_{ij})
  =&\;C_aT_R \frac{s_{aij}}{s_{ai}}\frac{4 \tilde{z}_a \tilde{z}_i \tilde{z}_j}{(1-\tilde{z}_j)^3}
  \left(1-2\cos^2\phi_{ai}^{jk}\right)\;.
  \end{split}
\end{equation}
We use the definition of the azimuthal angle in the soft-collinear approximation~\cite{Dulat:2018vuy}
\begin{equation}\label{eq:cosphi_def}
   4\,\tilde{z}_a\tilde{z}_i\cos^2\phi_{ai}^{jk}
   =\frac{(\tilde{z}_a s_{ij}-\tilde{z}_i s_{aj})^2}{
      s_{ai}\tilde{z}_j(s_{aj}+s_{ij})(\tilde{z}_a+\tilde{z}_i)}\;.
\end{equation}
For $s_{ai}\to 0$, this agrees with the definition of $\cos^2\phi_{a,j}^{i,k}$
in~\cite{Hoche:2017iem}. Away from the collinear limit, $\phi_{ai}^{jk}$ is not 
a physical angle, as $\cos^2\phi_{ai}^{jk}$ is not bounded by one. Equation~\eqref{eq:cosphi_def}
is constructed such that it reproduces the soft matrix element, hence the subtraction term 
$\mr{S}_{aq'}^{\rm(ds)}(\tilde{z},\Phi_{+1})$ provides a much better approximation of the
triple collinear and double soft matrix elements, leading to substantially smaller
real-emission contributions in Eq.~\eqref{eq:tcps_mc}.\footnote{The leading-order 
parton shower algorithm should implement spin correlations according to Eq.~\eqref{eq:tcsf_qqpa},
in order to achieve a consistent modified subtraction. While the parton shower we employ in
Sec.~\ref{sec:results} does not include these correlations yet, we note that their 
phenomenological impact is negligible except for dedicated observables, and we will 
therefore postpone their implementation to future work.}

In addition to the differential radiation pattern, the endpoint contributions need to be simulated.
This is achieved by extracting the $\mathcal{O}(1)$ contributions to the NLO splitting functions
that originate in the combination of the $-\delta(v)/\eps$ term in the series expansion 
in $v$, and the $\mc{O}(\eps)$ terms in the expansion of the differential forms of the subtraction
and matching terms. They are given by
\begin{equation}\label{eq:iterm_mc}
  \begin{split}
    \Delta\mr{I}_{aq'}^{\rm(ds)}(\tilde{z}_a,\tilde{z}_i,\tilde{z}_j)=&\;
    \tilde{\mr{I}}_{aq'}^{\rm(ds)}(\tilde{z}_a,\tilde{z}_i,\tilde{z}_j,\tilde{z}_a)-
    \tilde{\mc{I}}_{aq'}^{\rm(ds)}(\tilde{z}_a,\tilde{z}_i,\tilde{z}_j,\tilde{z}_a+\tilde{z}_i)\;,\\
  \end{split}
\end{equation}
where
\begin{equation}
  \begin{split}
    \tilde{\mr{I}}_{aq'}^{\rm(ds)}(\tilde{z}_a,\tilde{z}_i,\tilde{z}_j,\tilde{x})=&\;
    C_a T_R \left[\frac{2\tilde{z}_j}{1-\tilde{z}_j}
    \frac{2\,\tilde{z}_a\tilde{z}_i}{(\tilde{z}_a+\tilde{z}_i)^2}
      +\frac{2\tilde{z}_j}{1-\tilde{z}_j}
      \left(1-\frac{2\,\tilde{z}_a\tilde{z}_i}{(\tilde{z}_a+\tilde{z}_i)^2}\right)
      \log(\tilde{x}\,\tilde{z}_i\tilde{z}_j)\right]\;,\\
    \tilde{\mc{I}}_{aq'}^{\rm(ds)}(\tilde{z}_a,\tilde{z}_i,\tilde{z}_j,\tilde{x})=&\;
    C_a \frac{2\tilde{z}_j}{1-\tilde{z}_j}\log(\tilde{x}\,\tilde{z}_j)\,
    P_{gq}^{(0)}\Big(\frac{\tilde{z}_a}{\tilde{z}_a+\tilde{z}_i}\Big)\;.
  \end{split}
\end{equation}
The implementation of the endpoint contributions for $q\to\bar{q}$ transitions and 
the needed symmetry factors was discussed in~\cite{Hoche:2017iem} and remains unchanged.
The symmetry factors are reviewed in App.~\ref{sec:tagging}.

\section{Relation to the effective soft-gluon coupling}
\label{sec:cmw}
In this section we provide an intuitive explanation for the origin of the familiar
soft singular term $-20/9z$ in Eq.~\eqref{eq:p1_qqp} and~\eqref{eq:p1_qqp_ds}, 
and we explain how the two-loop cusp anomalous dimension emerges naturally 
upon integration over the final state phase space and summation over flavors.
Since the effective soft gluon coupling should be implemented as part of the soft-collinear
gluon radiation pattern~\cite{Dulat:2018vuy}, we conclude that a separation of the 
triple collinear splitting function into a double-soft component and a genuine 
triple collinear remainder yields an appropriate algorithm for parton shower evolution
at the next-to-leading order.

We first note that the endpoint contributions of the next-to-leading order 
splitting functions can be extracted by means of a series expansion 
of the scaled propagator virtuality, $v$
\begin{equation}
  \frac{1}{v^{1+\eps}}=-\frac{1}{\eps}\,\delta(v)
  +\sum_{i=0}^\infty\frac{\eps^n}{n!}\left(\frac{\log^n v}{v}\right)_+\;.
\end{equation}
When this term is combined with the $\mathcal{O}(\eps)$ contributions in the series
expansion of phase-space factors and splitting functions, it generates characteristic
logarithms, which contribute the leading transcendental terms to the anomalous dimensions.
In the triple collinear case we obtain
\begin{equation}\label{eq:tcps_tl_pole_1}
  \begin{split}
    \int{\rm d}\Phi_{+1}^{(F)}\frac{1}{v}=&\;
    \tilde{z}^{-\eps}
    \int_0^1{\rm d}\tilde{z}_j\,(\tilde{z}_j\tilde{z}_i)^{-\eps}
    \int_0^1{\rm d}v\,\frac{(1-v)^{-\eps}}{v^{1+\eps}}\;
    \frac{\Omega(1-2\eps)}{\Omega(2-2\eps)}
    \int_0^1{\rm d}\chi\,2(4\chi(1-\chi))^{-1/2-\eps}\\
    =&\;-\frac{\delta(v)}{\eps}
    \int_0^1{\rm d}\tilde{z}_j\,(\tilde{z}_j\tilde{z}_i\tilde{z})^{-\eps}\,
    \frac{\Omega(1-2\eps)}{\Omega(2-2\eps)}
    \int_0^1{\rm d}\chi\,2(4\chi(1-\chi))^{-1/2-\eps}
    +\ldots\;,
  \end{split}
\end{equation}
where the dots stand for plus distributions in $\ln^n v/v$. If the integrand
does not depend on the azimuthal angle variable $\chi$, we can simplify this to
\begin{equation}\label{eq:tcps_tl_pole_2}
  \begin{split}
    \int{\rm d}\Phi_{+1}^{(F)}\frac{1}{v}
    =&\;-\frac{\delta(v)}{\eps}
    \int_0^1{\rm d}\tilde{z}_j\,(\tilde{z}_j\tilde{z}_i\tilde{z})^{-\eps}\,
    +\ldots\;.
  \end{split}
\end{equation}
We can rewrite Eq.~\eqref{eq:tcps_tl_pole_2} as a convolution by measuring
$\tilde{z}$ and integrating over $\bar{\tau}=\tilde{z}/(1-\tilde{z}_j)$
\begin{equation}\label{eq:tcps_tl_pole_3}
  \begin{split}
    \int_0^1{\rm d}\bar{\tau}\,\int{\rm d}\Phi_{+1}^{(F)}\frac{1}{v}\,\delta(x-\tilde{z})
    =&\;-\frac{\delta(v)}{\eps}\,
    \int_x^1\frac{{\rm d}\bar{\tau}}{\bar{\tau}}\,\left(x\tilde{z}_i\left(1-\frac{x}{\bar{\tau}}\right)\right)^{-\eps}\,
    +\ldots\;.
  \end{split}
\end{equation}
Let us now consider the matching term in Eq.~\eqref{eq:twoloop_ffmatch}. It is given
by a similar convolution, but includes only the phase-space factors for the
production of the leading-order final state
\begin{equation}\label{eq:mtps_pole}
  \begin{split}
    \int_0^1{\rm d}\bar{\tau}\,\int{\rm d}\Phi_{+1}^{(F,J)}\frac{1}{v}\,\delta(x-\tilde{z})
    =&\;-\frac{\delta(v)}{\eps}\,\int_x^1\frac{{\rm d}\bar{\tau}}{\bar{\tau}}\,
    \left(\left(1-\frac{x}{\bar{\tau}}\right)\frac{x}{\bar{\tau}}\right)^{-\eps}\,+\ldots\;.
  \end{split}
\end{equation}
Combining the complete phase-space integral and the integral
needed for the matching term leads to
\begin{equation}
  \begin{split}
    &\int_0^1{\rm d}\bar{\tau}\,\left(\int{\rm d}\Phi_{+1}^{(F)}
    -2\int{\rm d}\Phi_{+1}^{(F,J)}\right)\,\frac{1}{v}\,\delta(x-\tilde{z})\\
    &\qquad=-\frac{\delta(v)}{\eps}\,\int_x^1\frac{{\rm d}\bar{\tau}}{\bar{\tau}}\,
    \bigg[\,1-\eps\ln\left(\bar{\tau}(1-\bar{\tau})\right)+\eps\ln\left(1-\frac{x}{\bar{\tau}}\right)
    +\mathcal{O}(\eps^2)\,\bigg]\,+\ldots\;.
  \end{split}
\end{equation}
In the double soft limit, $x/\bar{\tau}\to 0$, and we nearly recover the standard double-collinear
phase-space integral. Applying this to the approximate splitting function in 
Eq.~\eqref{eq:tcsf_qqpa}, we can reconstruct the leading soft enhanced term as
\begin{equation}
    \begin{split}
    &\int_0^1{\rm d}\bar{\tau}\,\left(\int{\rm d}\Phi_{+1}^{(F,\rm ds)}
    P_{ag}^{\rm(s)}\left(\frac{x}{\bar{\tau}}\right)P_{gq'}^{(0)}(\bar{\tau},\eps)
    -2\int{\rm d}\Phi_{+1}^{(F,J,\rm ds)}
    P_{ag}^{\rm(s)}\left(\frac{x}{\bar{\tau}}\right)P_{gq'}^{(0)}(\bar{\tau},0)\right)\,
    \frac{1}{v}\,\delta(x-\tilde{z})\\
    &\quad=\mathcal{O}\Big(\frac{1}{\eps}\Big)+\delta(v)\,
    \int_x^1\frac{{\rm d}\bar{\tau}}{\bar{\tau}}\,P_{ag}^{\rm(s)}\left(\frac{x}{\bar{\tau}}\right)
    \left[P_{gq'}^{(0)}(\bar{\tau},0)\big(\ln(\bar{\tau}(1-\bar{\tau}))+1\big)
    -P_{gq'}^{(0)}(\bar{\tau},\eps)\right]\,+\ldots+\mathcal{O}(\eps)\;,
    \end{split}
\end{equation}
where the dots stand for contributions that are finite in $x$, and for plus distributions
in $\ln^n v/v$. The function $P_{ag}^{\rm(s)}(z)=2C_a(1-z)/z$ is the soft-collinear 
splitting kernel for the transition $a\to g$. Its leading term is given by $2C_a/z$,
such that we can extract the leading term in $1/x$ of the finite remainder as
\begin{equation}\label{eq:cmw_eq}
    \delta(v)\,\frac{2C_a}{x}\,T_R\,\int_x^1{\rm d}\bar{\tau}\,
    \Big[(1-2\bar{\tau}(1-\bar{\tau}))\ln(\bar{\tau}(1-\bar{\tau}))+2\bar{\tau}(1-\bar{\tau})\Big]
    \overset{x\to0}{\longrightarrow}
    \delta(v)\,\frac{2C_a}{x}\,T_R\left(-\frac{10}{9}+\mathcal{O}(x)\right)\;.
\end{equation}
The relation to the CMW scheme is now manifest: The leading term on the right-hand side of
Eq.~\eqref{eq:cmw_eq} is simply the contribution from the production of a single quark pair
to the $n_f$ term in the two-loop cusp anomalous 
dimension~\cite{Kodaira:1981nh,Davies:1984hs,Davies:1984sp,Catani:1988vd}.
The parentheses do not evaluate to a constant, because we explicitly consider 
resolved partons, i.e.\ we implement the measurement $\delta(x-\tilde{z})$.

In summary, we find that the above contribution from the soft-collinear 
splitting function correctly reproduces the expected finite remainders
in the soft limit, and therefore its simulation in fully differential form
induces the conventional rescaling of the soft-gluon coupling~\cite{Catani:1990rr} 
upon integration over $x$. As outlined in Sec.~\ref{sec:basic}, 
it is therefore appropriate to implement the corresponding endpoints
as part of the soft gluon radiator function~\cite{Dulat:2018vuy}.

We finally note that the above calculation serves only to make the origin
of the soft gluon coupling explicit. We do not explicitly implement
the resolved parton evolution in our numerical simulations. Instead, following
the derivation in~\cite{Jadach:2003bu,Hoche:2017iem}, we obtain equivalent results
from unconstrained parton evolution including tagging factors,
which allows us to implement the soft physical coupling using the technique
described in~\cite{Dulat:2018vuy}. This method is reviewed in App.~\ref{sec:tagging} 
As a direct consequence, we are not restricted 
to interpreting the $q\to q q'\bar{q}'$ transitions that we implement 
as a contribution solely to the NLO $q\to q'$ splitting function. The same splitting
also contributes to the $n_f$ part of the $q\to q$ splitting function.
This is achieved by summing over all possible ways to tag the 
final-state partons. We have thus presented a generic technique to implement
all real-emission type $n_f$ contributions to the next-to-leading order
splitting functions, as well as the corresponding $C_F-C_A/2$ interference terms.

\section{Numerical results}
\label{sec:results}

\begin{figure}[t]
  \subfigure{
    \begin{minipage}{0.475\textwidth}
      \begin{center}
        \includegraphics[scale=0.7]{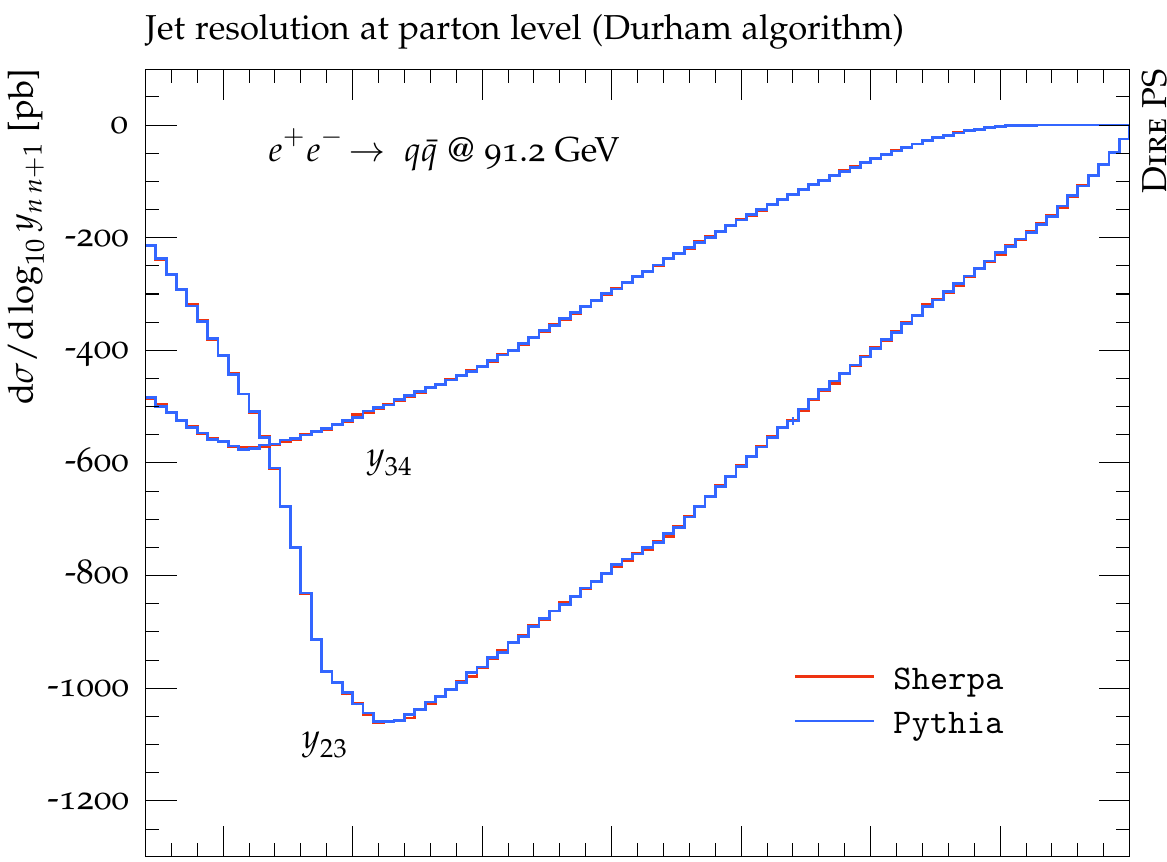}\\[-0.75mm]
        \includegraphics[scale=0.7]{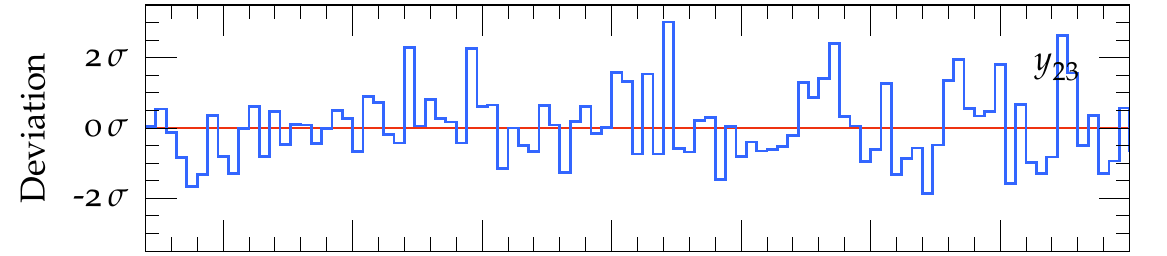}\\[-0.75mm]
        \includegraphics[scale=0.7]{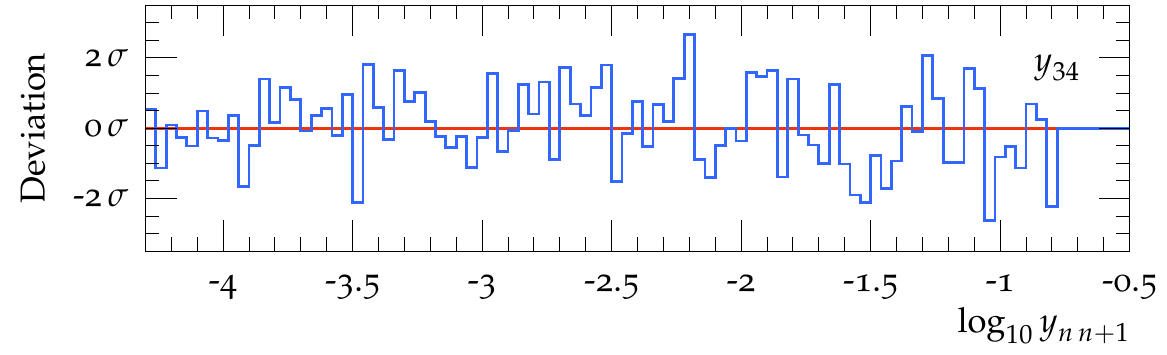}
      \end{center}
    \end{minipage}
    \label{fig:validation_ff}}
  \subfigure{
    \begin{minipage}{0.475\textwidth}
      \begin{center}
        \includegraphics[scale=0.7]{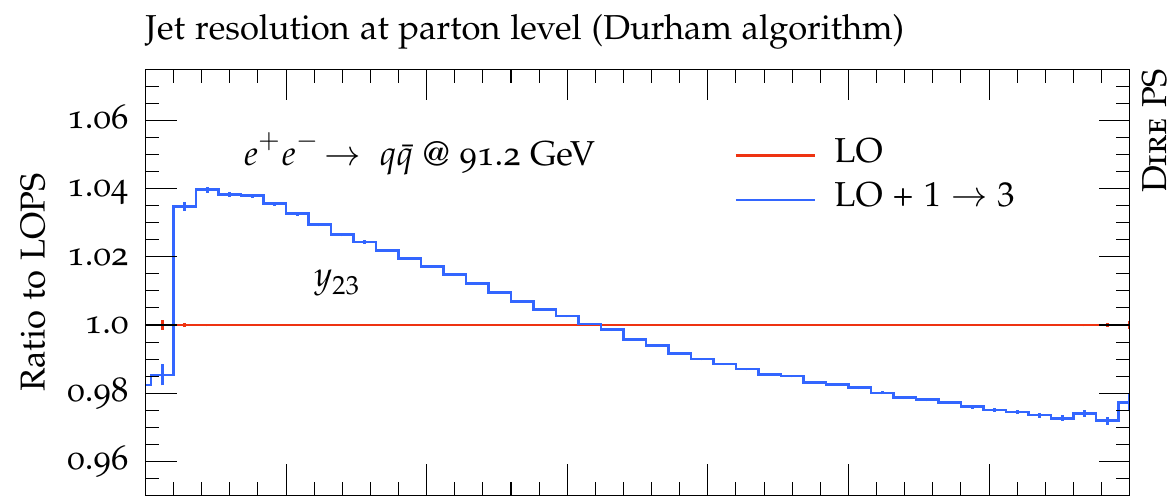}\\[-0.5mm]
        \includegraphics[scale=0.7]{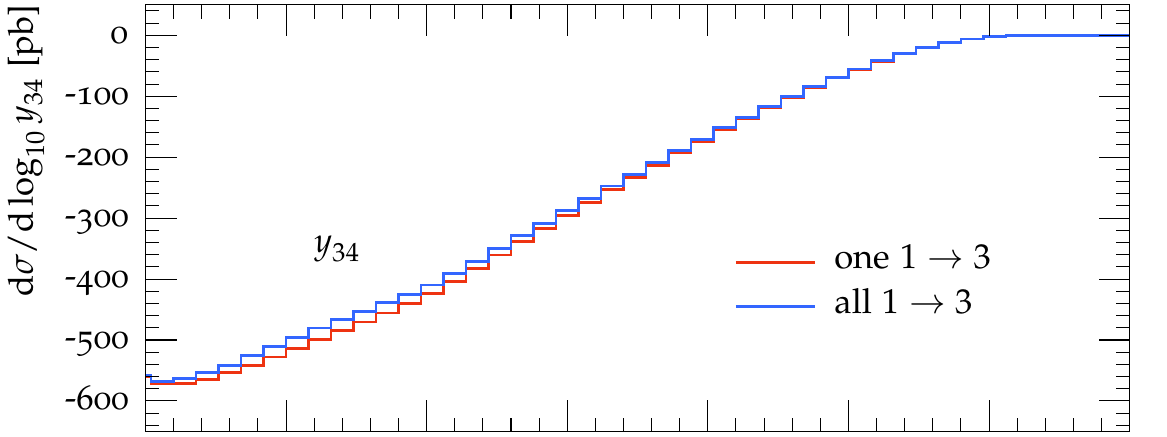}\\[-0.5mm]
        \includegraphics[scale=0.7]{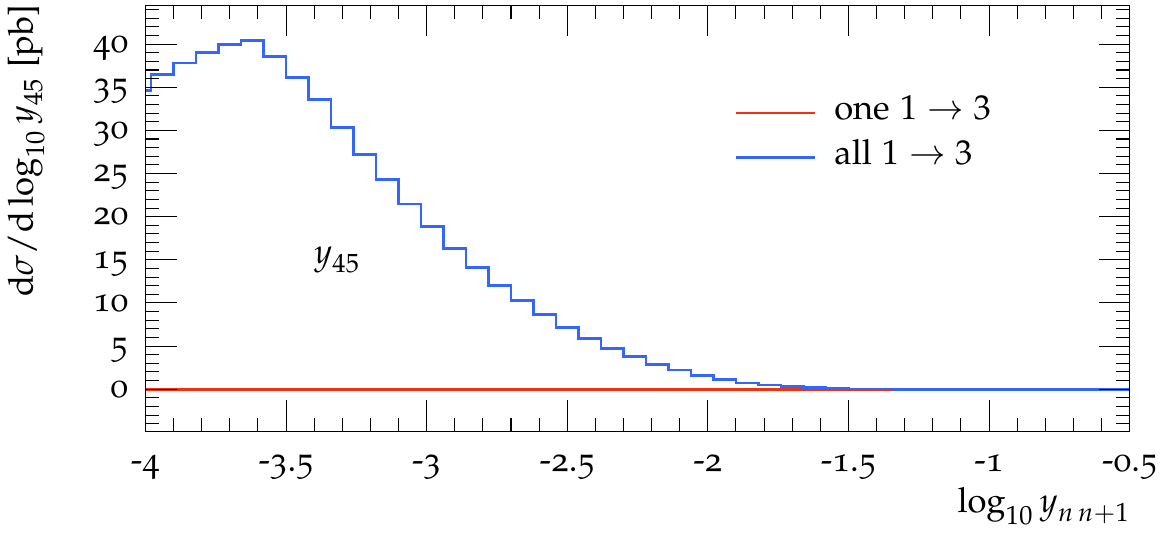}
      \end{center}
    \end{minipage}
    \label{fig:impact_ff}}
  \caption{Durham $k_T$-jet rates in $e^+e^-\to$ hadrons at LEP.
  Left: Validation of the simulation of soft-subtracted triple-collinear parton splittings.
  Right: Impact of the soft-subtracted triple-collinear simulation. 
  The top panel shows the ratio between the leading-order result and the
  leading-order simulation including soft-subtracted triple-collinear branchings. 
  The middle and bottom panels show a comparison between the simulation
  of up to one soft-subtracted triple-collinear splitting and arbitrarily many 
  (both not including the leading-order result).
  \label{fig:validation_impact}}
\end{figure}

In this section we present the first application of our algorithm to the process
$e^+e^-\to$ hadrons at LEP energies. We implement a computation of the soft-subtracted
$q\to qq'\bar{q}'$ and $q\to qq\bar{q}$ triple collinear splitting functions into 
the \Dire parton showers, which provide two entirely independent codes within
the event generation frameworks \Pythia~\cite{Sjostrand:1985xi,Sjostrand:2014zea}
and \Sherpa~\cite{Gleisberg:2008ta,Sherpa:2019gpd}.
We employ the CT10nlo PDF set~\cite{Lai:2010vv}, and use the corresponding form
of the strong coupling. Following standard practice, we implement the CMW scheme
through a rescaling of the soft gluon coupling by $1+\alpha_s(t)/(2\pi) K$, where
$K=(67/18-\pi^2/6)\,C_A-10/9\,T_R\,n_f$~\cite{Catani:1990rr}.
The implementation of this term in fully differential form has been discussed 
in~\cite{Dulat:2018vuy}.

The left side in Fig.~\ref{fig:validation_impact} shows a comparison
between the \textsc{Dire+Sherpa} and \textsc{Dire+Pythia} predictions 
for the soft-subtracted triple-collinear $q\to qq'\bar{q}'$ splittings, 
when considering only a single branching. The lower panel shows the deviation 
of the two results, normalized bin-wise to the statistical uncertainty. 
We find perfect agreement, suggesting that no technical problems are present. 
A single $1\to 3$ branching populates both the $2\to3$ jet rate, $y_{23}$, 
and the $3\to4$ jet rate, $y_{34}$. The $3\to 4$ jet rate is entirely given 
by the $R-S$ contribution in Eq.~\eqref{eq:tcps_mc}, while the $2\to 3$ jet rate 
also receives contributions from the $I - \mathcal{I}$ term. The contributions 
from the soft-subtracted triple-collinear branchings are negative, as anticipated 
based on Eq.~\eqref{eq:collAnomDim}, and they are of similar size for both rates. 
To be consistent with the renormalization group evolution of the strong coupling, 
we only produce b-quarks if the shower evolution variable is above the quark mass, 
$t>m_b^2$. The corresponding threshold effects can be seen close to
$\log_{10}(4.75^2/91.2^2) = -2.6$. A similar effect for the charm quark is not visible,
since the threshold at $\log_{10}(1.3^2/91.2^2) = -3.7$ is too close to the parton-shower 
cutoff placed at 1~GeV.

The right side of Fig.~\ref{fig:validation_impact} shows the phenomenological impact
of the soft-subtracted triple-collinear branchings. The upper panel displays 
the ratio between the pure leading-order parton evolution and the LO + $1\to 3$ evolution,
indicating a difference of up to $4\%$ in the $2\to 3$ jet rate. Compared to the 
triple-collinear $q\to q'$ and $q\to \bar{q}$  corrections presented in \cite{Hoche:2017iem},
we find larger effects, since we not only consider the contribution to the identified 
final state, but the sum over all ways to tag the $qq'\bar{q}'$ or $qq\bar{q}$ final-state
(cf.\ the last paragraph of Sec.~\ref{sec:cmw}). Allowing for multiple $1\to 3$ branchings
has a marginal effect on the $3\to 4$ jet rate, and adds a very small correction 
to the $4\to 5$ jet rate. These are shown in the middle and bottom panels of Fig.~\ref{fig:impact_ff}.

\section{Conclusions}
\label{sec:conclusion}
This note introduced a method for the consistent combination of triple-collinear 
and double-soft corrections to parton evolution at leading-order by means of subtraction 
at the integrand level. We argue that a subtraction technique is the most appropriate
method for addressing the soft-collinear overlap, as it allows to cleanly separate 
the integrands into soft enhanced and soft finite contributions. It is also supported
by the fact that the effective soft-gluon coupling generated by the radiative corrections
in the triple collinear limit can be obtained by including double soft corrections alone.

In our algorithm, all higher-order corrections are embedded in the parton shower 
in fully differential form, using the appropriate transition matrix elements computed
in dimensional regularization and the $\overline{\rm MS}$ scheme. The method recovers
known analytic results, such as the $n_f$ contribution to the two-loop cusp anomalous 
dimension. While we explicitly considered only the special case of quark pair emission from quarks, 
we note that other triple collinear splitting functions can be treated in the same manner.

We have implemented our new method into two independent Monte-Carlo programs 
in the general-purpose event generators \Pythia and \Sherpa for the case of 
$q\to q'\bar{q}'$ and $q\to qq\bar{q}$ transitions, proving the feasibility 
of the algorithmic considerations for numerical studies. Overall, the impact 
of the genuine triple-collinear corrections to the parton cascade is small 
for standard observables -- provided that the leading-order shower correctly 
reproduces the radiation pattern at $\mathcal{O}(\alpha_s^2)$ in ordered 
phase-space regions. This supports previous findings that the main effect 
of the $\mathcal{O}(\alpha_s^2)$ corrections is to reduce the uncertainties 
present in the leading-order simulation.

\section{Acknowledgments}
We thank Joshua Isaacson for comments on the manuscript. 
This note was supported by funding from the Swedish Research Council,
contract numbers 2016-05996 and 2020-04303,
and by the Fermi National Accelerator Laboratory (Fermilab),
a U.S. Department of Energy, Office of Science, HEP User Facility.
Fermilab is managed by Fermi Research Alliance, LLC (FRA),
acting under Contract No. DE--AC02--07CH11359. 
We further acknowledge funding from the European Union’s Horizon 2020
research and innovation program as part of the 
Marie Sk{\l}odowska-Curie Innovative Training Network
MCnetITN3 (grant agreement no. 722104).

\appendix
\section{Implementation of symmetry factors}
\label{sec:tagging}
In this appendix we review the techniques of~\cite{Hoche:2017iem}, which
allow to consistently implement final-state parton evolution by means of 
the Sudakov factor
\begin{equation}\label{eq:sudakov}
  \Delta_a(t_0,t)=\exp\bigg\{-\int_{t_0}^{t}\frac{{\rm d} \bar{t}}{\bar{t}}
  \sum_{c=q,g} \int_{z_-}^{z_+}{\rm d}z\,z\,\frac{\alpha_s}{2\pi}P_{ac}(z)\bigg\}\;.
\end{equation}
Here, $z_-$ and $z_+$ stand for the lower and upper integration limit on the 
splitting variable, and we highlight the additional factor $z$ multiplying 
the splitting function, which corresponds to using the momentum sum rule in order 
to satisfy the unitarity constraint on the parton shower evolution~\cite{Jadach:2003bu}.
At the next-to-leading order, we are required to multiply the real-emission
correction to the splitting functions in Eq.~\eqref{eq:tcps_mc} by two
such splitting variables in order to obtain the correct sum over flavors.
This can be interpreted as an identification, or `tagging' of the identified
parton whose evolution is considered. More precisely, we find
\begin{equation}\label{eq:nlokernels_symmetry}
  \begin{split}
    &\sum_{b=q,g}\int_{0}^{1-\eps} {\rm d}z_1
    \int_{0}^{1-\eps} {\rm d}z_2\,\frac{z_1\,z_2}{1-z_1}\,
    \Theta(1-z_1-z_2)\,P_{a\to ab\bar{b}}(z_1,z_2,\ldots)\\
    &\qquad=\sum_{b=q,g}\int_{\eps}^{1-\eps} {\rm d}z_1\,
    \int_{\eps}^{1-z_1} {\rm d}z_2\,S_{ab\bar{b}}\,P_{a\to ab\bar{b}}(z_1,z_2,\ldots)
    +\mc{O}(\eps)\;,\\
    &\sum_{\substack{b=q,g\\b\neq a}}\int_{0}^{1-\eps} {\rm d}z_1
    \int_{0}^{1-\eps} {\rm d}z_2\,\frac{z_1\,z_2}{1-z_1}\,\Theta(1-z_1-z_2)\,
    \Big(P_{a\to ba\bar{b}}(z_1,z_2,\ldots)+P_{a\to b\bar{b}a}(z_1,z_2,\ldots)\Big)\\
    &\qquad=\sum_{\substack{b=q,g\\b\neq a}}\int_{\eps}^{1-\eps} {\rm d}z_1\,
    \int_{\eps}^{1-z_1} {\rm d}z_2\,S_{ab\bar{b}}\,P_{a\to ba\bar{b}}(z_1,z_2,\ldots)
    +\mc{O}(\eps)\;,
  \end{split}
\end{equation}
where $S_{ab\bar{b}}=1/(\prod_{c=q,g}n_c!)$, with $n_c$ the number of partons
of type $c$, is the usual symmetry factor for the final-state $ab\bar{b}$.
Thus, the subtracted real-emission corrections in Eq.~\eqref{eq:tcps_mc}
should be multiplied by the final-state symmetry factor (or tagging factor)
$S^{(F)}=z_1 z_2/(1-z_1)$.

\bibliography{journal}

\end{document}